# Innovation in Scholarly Communication: Vision and Projects from High-Energy Physics[1]

*Rolf-Dieter Heuer, Annette Holtkamp*
*Deutsches Elektronen-Synchrotron DESY*
*Notkestraße 85, D-22607 Hamburg, Germany*

*Salvatore Mele*
*CERN, European Organization for Nuclear Research*
*CH1211, Genève 23, Switzerland*

## Abstract

Having always been at the forefront of information management and open access, High-Energy Physics (HEP) proves to be an ideal test-bed for innovations in scholarly communication including new information and communication technologies. Three selected topics of scholarly communication in High-Energy Physics are presented here: A new open access business model, SCOAP$^3$, a world-wide sponsoring consortium for peer-reviewed HEP literature; the design, development and deployment of an e-infrastructure for information management; and the emerging debate on long-term preservation, re-use and (open) access to HEP data.

## 1. Preamble

Research in High-Energy Physics (HEP), also called Particle Physics, is motivated by the goal of attaining a fundamental description of the laws of physics, such as explaining the origin of mass, and understanding the dark matter in the universe. Although fundamentally driven by the quest for knowledge, the ensuing research is performed at the edge of what is feasible technologically and hence drives the development of technology in many areas. The knowledge gained from studying the microcosm of particle collisions at the highest energies ever attained provides also insight into the early universe and its development since its creation. To further this understanding, experimental particle physicists build the largest instruments ever to reach energy densities close to the Big Bang, teaming up in collaborations of up to several thousands of scientists. At the same time, theoretical particle physicists, who represent the other half of the community, build hypotheses and theories to accommodate and predict experimental findings.

---

[1]Based on a keynote talk by Rolf-Dieter Heuer at APE2008, International Conference on "Academic Publishing in Europe", Berlin, January 22$^{nd}$-23$^{rd}$ 2008.

HEP experimental research takes place in international accelerator research centres in Europe, such as the European Organization for Nuclear Research (CERN) in Geneva or the Deutsches Elektronen-Synchrotron (DESY) in Hamburg; in the United States mainly at the Stanford Linear Accelerator Center (SLAC) in California and the Fermi National Accelerator Laboratory (Fermilab) in Illinois; and in Japan at the High Energy Accelerator Research Organization (KEK) in Tsukuba. Canada, China, and Italy host other laboratories. HEP theoretical research takes place in hundreds of universities and institutes worldwide, which also host experimental teams building parts of the large detectors used at the large accelerator laboratories and analyzing the data collected with them.

With the start-up of CERN's Large Hadron Collider (LHC) in 2008 and preparations for the International Linear Collider (ILC) in full swing, we expect revolutionary results explaining the origin of matter, unravelling the nature of dark matter and providing glimpses of extra spatial dimensions or grand unification of forces. Any of these insights would dramatically change our view of the world.

The LHC will collide protons 40 million times a second and reproduce the conditions at the origin of the universe. These collisions will be observed by large detectors, up to the size of a five storey building, crammed with electronic sensors: think a 100MegaPixel digital camera taking 40 million pictures a second.

This contribution will not describe the exciting intellectual, scientific and technological endeavour of particle physics. Rather it will attempt to describe some solutions that, on the long wave of its track record in international collaboration, HEP has proposed and implemented for its infrastructure in scholarly communication. Although this is a very discipline-specific approach to the present evolution of scholarly communication, it might be of interest for an audience learned in academic publishing.

## 2. Introduction

Progress in information and communication technologies is driving evolving needs and profound changes in scholarly communication. Scientists, and not only HEP scholars, have come to expect:
- easy and unrestricted access to comprehensive scientific information in their field and cognate disciplines;
- state-of-the art information tools to optimize their research workflow, with powerful discovery tools and limited noise;
- quality assurance, at the intersection of three centuries of tradition in peer-review, but with a twist of XXI$^{st}$ century communication for immediacy of feedback and dissemination.

At the same time, these desires have to be balanced against budget efficiency and optimization of resources for research. HEP has been proposing solutions to these needs since decades, as described in Section 3, while HEP *ante-litteram* open access tradition, which dates back half a century, is discussed in Section 4.

With the intention of informing, and possibly inspiring, the ongoing debates in the wider arena of innovation in scholarly communication, and its intersection with academic publishing in Europe and beyond, this contribution discusses the vision of HEP along three axes of innovation: a new open access business model (Section 5); the design, development and deployment of an e-infrastructure for information management, a next-generation repository (Section 6); the emerging debate on long-term preservation, re-use and (open) access to HEP data (Section 7).

## 3. Scholarly communication in HEP

To set the scene, it is useful to quote five numbers and a concept. The five numbers, which parameterize scholarly communication in HEP, are:
- 20'000, a lower limit to the number of active HEP scholars;
- 10'000, an upper limit to the yearly number of HEP articles;
- 80%, the fraction of these articles produced by theoretical physicists working alone or in small teams in hundreds of universities worldwide;
- 20%, the fraction of these articles authored by large collaborations of experimental physicists, centered around half a dozen international laboratories[2];
- 50:50, the ratio of active experimental and theoretical HEP scholars.

The concept is the one of *preprint culture,* which is discussed in the following.

The *preprint culture* in HEP pioneered the free distribution of scientific results. For decades, theoretical physicists and scientific collaborations, eager to disseminate their results in a way faster than the distribution of scholarly publications, printed and mailed hundreds, even thousands, of copies of their manuscripts at the same time as submitting to peer-reviewed journals[3]. While assuring the broadest possible dissemination of scientific results, albeit privileging scientists working in affluent institutions, this *ante-litteram* form of "author-pays" or rather "institute-pays" open access implied non-negligible financial liabilities for research centres: as an example, in the '90s DESY used to spend about 1 Million DM (500'000 €, not corrected for inflation) a year for the production and mailing of hard-copies of these preprints. CERN used to spend about twice as much.

From the very beginning of the preprint culture, HEP libraries classified preprints received from all over the world and regularly distributed worldwide information about the latest accessions. This was a pioneering instance of a literature database later evolving in email alerts, and the only way for HEP scholars to keep track of the fast-paced advancement of the field. As an example, the DESY library started in 1963 to publish the biweekly "HEP Index". It contained bibliographic information about the newest HEP preprints, journal articles and conference proceedings, and in addition it grouped these items according to standardized keywords.

---

[2]S. Mele *et al.,* Journal of High Energy Physics 12 (2006) S01, arXiv:cs.DL/0611130
[3]L. Goldschmidt-Clermont, 1965, *Communication Patterns in High-Energy Physics,*
http://eprints.rclis.org/archive/00000445/02/communication_patterns.pdf

In this scene, three revolutions mark the advances in scholarly communication in HEP, with repercussions in the contemporary innovations affecting other disciplines.

**1974, information technology meets (HEP) libraries.** The SPIRES database, the first grey-literature electronic catalogue, saw the light at SLAC[4]. Shortly thereafter the SLAC and DESY libraries joined forces to cover the complete HEP literature including preprints, reports, journal articles, theses, conference talks and books. In 1985, the database contained already more than 140,000 records. It now contains metadata for about 760,000 HEP articles, including links to full-text, standardized keywords, publication notes. It offers additional tools like citation analysis and is interlinked with other databases containing information on conferences, experiments, authors and institutions.

**1991, the first repository.** arXiv, the archetypal repository, was conceived in 1991 by Paul Ginsparg[5], then at the Los Alamos National Laboratory in New Mexico, and is now hosted at Cornell University in New York. It evolved the four-decade old preprint culture into an electronic system, offering all scholars a level playing-field from which to access and disseminate information. Today arXiv has grown outside the field of HEP, becoming the reference repository for many diverse disciplines beyond physics, from mathematics to some areas of biology. It contains about 450'000 full-text preprints, receiving about 5'000 new articles each month.

**1991, the web is woven.** The early history of the web and its invention at CERN in 1991 by Sir Tim Berners-Lee are today household stories[6]. What is less known is that the first web server outside Europe was installed at SLAC to provide access to the SPIRES database, which had then the honour to be the first database on the web[7]. In Summer '92, SPIRES linked to the arXiv for full-texts, starting a close partnership, and bringing preprints on the web, accessible through a detailed indexing including reference to the ensuing published versions.

Even now, in the era of electronic journals, the preprint culture lives on in HEP thanks to the speed and ease of access. Journals have to a large extent lost their century-old role as vehicles of scholarly communication. However, at the same time, they continue to play a crucial part in the HEP community. Evaluation of research institutes and researchers, especially young ones at the beginning of their career, is largely based on publications in prestigious peer-reviewed journals. The main role of journals in HEP is perceived as the one of "keeper-of-the-records", by guaranteeing a high-quality peer-review process. The HEP community needs high-quality journals as its "interface with officialdom".

## 4. HEP and Open Access

Thanks to decades of preprint culture, today, in open-access speak, HEP could be defined as an almost entirely "green" discipline, where authors self-archive their research results. Posting an article on arXiv even before submitting it to a journal is common practice. Even revised versions incorporating the changes due to the peer-review process are routinely uploaded. It is interesting to remark that this comes without mandates and without debates: very few HEP scientists would not take advantage of the formidable opportunities offered by the discipline repository of the field, and the linked discovery and citation-analysis tools.

---

[4] L. Addis, 2002, http://www.slac.stanford.edu/spires/papers/history.html (last visited May 1st, 2008);
P. A. Kreitz and T. C. Brooks, Sci. Tech. Libraries 24 (2003) 153, arXiv:physics/0309027
[5] P. Ginsparg, Computers in Physics 8 (1994) 390
[6] T. Berners-Lee, Weaving the Web, HarperCollins, San Francisco (1999)
[7] P. Kunz et al, http://www.slac.stanford.edu/history/earlyweb/history.shtml (last visited May 1st, 2008)

The synergy between HEP and open access[8] extends beyond preprints. In 1997, HEP launched one of the first peer-reviewed open access journals: the *Journal of High Energy Physics* (JHEP), courtesy of the International School of Advanced Studies (SISSA) in Trieste. It was followed in 1998 by *Physical Review Special Topics Accelerators and Beams*, published by the American Physical Society, and the *New Journal of Physics*, published by the Institute of Physics Publishing, which carries HEP content in a broader spectrum of content covering many branches of physics.

After preprints, arXiv and the web, open access journals appear to be the next logical step in the natural evolution of HEP scholarly communication. It is remarkable that in this field, this call for open access journals is not only originating from librarians frustrated by spiralling subscription costs and shrinking budget, but is a solid pillar of the scientific community. At the beginning of 2007, the four LHC Collaborations ATLAS, CMS, ALICE and LHCb, counting over 5'000 scientists, declared: *"We, […] strongly encourage the usage of electronic publishing methods for [our] publications and support the principles of open access Publishing, which includes granting free access of our publications to all. Furthermore, we encourage all [our] members to publish papers in easily accessible journals, following the principles of the open access paradigm."*[9]

The Helmholtz Alliance 'Physics at the Terascale', a German network comprising theorists, experimentalists, computer and accelerator scientists from 17 universities, two Helmholtz institutes and one Max Planck institute, issued a similar statement in January 2008: *"The Strategic Helmholtz Alliance 'Physics at the Terascale' fully supports the goal […] of free and unrestricted electronic access to peer-reviewed journal literature in particle physics. [It] will benefit scientists, authors, funding agencies and publishers alike. Unrestricted access to published scientific results is essential for wide dissemination and efficient usage of scientific knowledge. [It invites its members to] raise awareness on open-access publishing in their communities and […] to publish in open-access journals."*[10]

In order to meet these goals the SCOAP[3] initiative, subject of the next section, was started.

## 5. The SCOAP[3] Initiative

SCOAP[3], the Sponsoring Consortium for Open Access Publishing in Particle Physics, is an initiative that aims to convert to open access the HEP peer-reviewed literature in a way transparent to authors[11]. Its business model originates from a debate involving the scientific community, libraries and publishers[12]. The essence of this model is the formation of a consortium to sponsor HEP publications and make them open access by redirecting funds that are currently used for subscriptions to HEP journals. Today, libraries (or the funding bodies behind them) buy journal subscriptions to support the peer-review service and to allow their users to read articles, even though these mostly access their information by reading preprints on arXiv. The SCOAP$^3$ vision for tomorrow is that funding bodies and libraries worldwide federate in a consortium that will pay centrally for the peer-review service and that journal articles will be free to read for everyone. This evolution of the current "author-pays" models for open access attempts to make the transition to open access transparent for authors, by removing any barriers.

---

[8] There are many definitions and flavours of open access. For the sake of clarity, in this contribution we will assume something along the lines of "grant anybody anywhere and anytime access to the (peer-reviewed) results of (publicly-funded) research"… in HEP, of course.
[9] S. Bianco *et al.*, Report of the SCOAP[3] Working Party, http://www.scoap3.org/files/Scoap3WPReport.pdf
[10] http://www.terascale.de/news/scoap3_initiative (last visited May 1st, 2008)

SCOAP$^3$ will sponsor articles through a tendering procedure with publishers of high-quality journals. It has therefore the potential to contain the overall cost of journal publishing by linking price with quality and injecting competition into the market.

In practice, the open access transition will be facilitated by the fact that the large majority of HEP articles are published in just six peer-reviewed journals from four publishers[13], as presented in Figure 1.

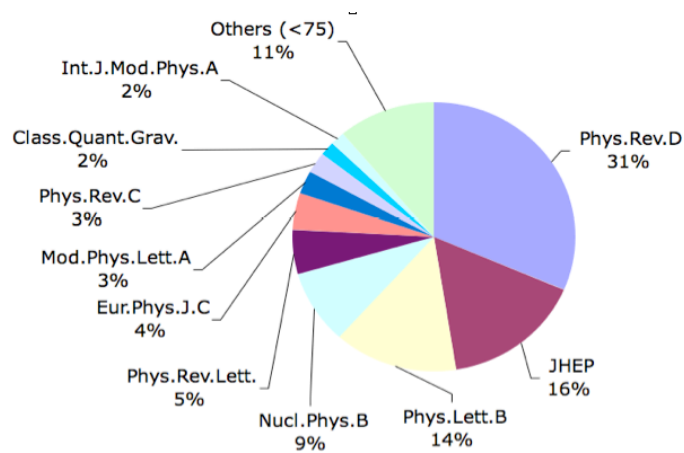

*Figure 1. Journals favoured by HEP scientists in 2006. Data from the SPIRES database. Journals that attracted less than 75 HEP articles are grouped in the slice named "Others".*

Five of those six journals carry a majority of HEP content. These are *Physical Review D* (published by the American Physical Society), *Physics Letters B* and *Nuclear Physics B* (Elsevier), *Journal of High Energy Physics* (SISSA/IOP) and the *European Physical Journal C* (Springer). The aim of the SCOAP$^3$ model is to assist publishers to convert these "core" HEP

---

[11] S. Bianco *et al.*, Report of the SCOAP$^3$ Working Party, http://www.scoap3.org/files/Scoap3WPReport.pdf; http://scoap3.org
[12] R. Voss *et al.*, Report of the Task Force on Open Access Publishing in Particle Physics, http://www.scoap3.org/files/cer-002632247.pdf
[13] S. Mele *et al.*, Journal of High Energy Physics 12 (2006) S01, arXiv:cs.DL/0611130

journals entirely to open access and it is expected that the vast majority of the SCOAP³ budget will be spent to achieve this target. The sixth journal, *Physical Review Letters* (American Physical Society), is a "broadband" journal that carries only a small fraction (10%) of HEP content; it is the aim of SCOAP³ to sponsor the conversion to open access of this journal fraction. The same approach can be extended to another "broadband" journal popular with HEP instrumentation articles: *Nuclear Instruments and Methods in Physics Research A* (Elsevier) with about 25% HEP content. Of course, the SCOAP³ model is open to any other, present or future, high-quality journals carrying HEP content. This will ensure a dynamic market with healthy competition and a broader choice.

The price of an electronic journal is mainly driven by the costs of running the peer-review system and editorial processing. Most publishers quote a price in the range of 1'000–2'000€ per published article. On this basis, given that the total number of HEP publications in high-quality journals is between 5'000 and 10'000, according to how one defines HEP and its overlap with cognate disciplines, the annual SCOAP³ budget for the transition of HEP publishing to open access would amount to a maximum of 10 Million Euros per year.

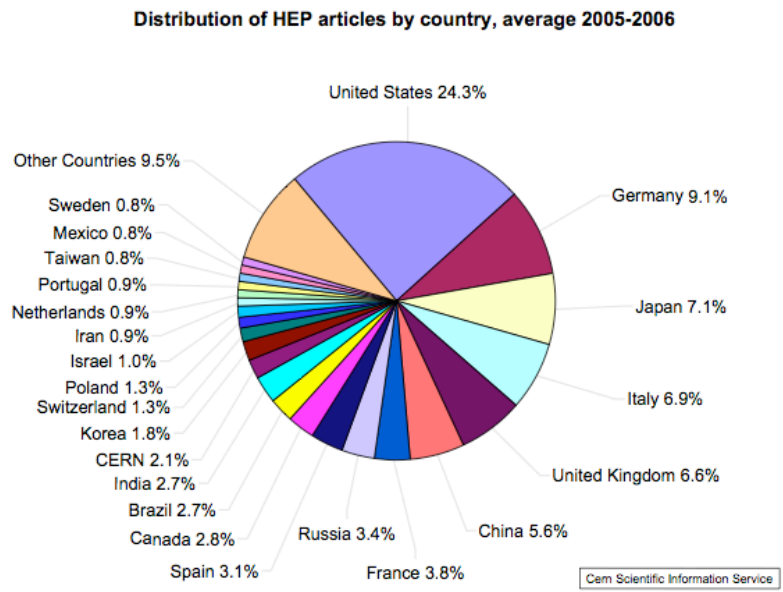

*Figure 2. Contributions by country to the HEP scientific literature published in the largest journals in the field. Co-authorship is taken into account on a pro-rata basis, assigning fractions of each article to the countries in which the authors are affiliated. This study is based on over 11'000 articles published in the years 2005 and 2006. Countries with individual contributions less than 0.8% are aggregated in the "Other countries" category[14].*

The costs of SCOAP³ will be distributed among all countries according to a fair-share model based on the distribution of HEP articles per country, as shown in Figure 2. In practice, this is an evolution of the "author-pays" concept: countries will be asked to contribute to SCOAP$^3$, whose ultimate targets are open access and peer-review, according to their use of the latter, a.k.a. their scientific productivity. To cover publications from scientists from countries that cannot be reasonably expected to make a contribution to the consortium at this time, an allowance of not more than 10% of the SCOAP$^3$ budget is foreseen.

---

[14] J. Krause *et al.*, http://www.scoap3.org/files/cer-002691702.pdf

The SCOAP$^3$ initiative fits in the European-wide debate on the access to results of scientific research. The Council of the European Union, in the conclusions of its 2832$^{nd}$ Competitiveness Council, recognized "the strategic importance for Europe's scientific development of current initiatives to develop sustainable models for open access [...]", underlining "the importance of effective collaboration between different actors, including funding agencies, researchers, research institutions and scientific publishers, in relation to access [...to] scientific publications". These principles are precisely the pillars of the SCOAP$^3$ model. Finally, it "invite[d] Member States to enhance the co-ordination between [...] large research institutions and funding bodies on access [...] policies and practices"[15].

It appears at first glance to be a formidable enterprise to organize a worldwide consortium of research institutes, libraries and funding bodies that cooperates with publishers in converting the most important HEP journals to OA. At the same time, HEP is used to international collaborations on a much bigger scale. As an example, the ATLAS experiment, one of the four detectors at the LHC, has been built over more than a decade by about 50 funding agencies on a total budget of 400M€ (excluding person-power), placing about 1000 industrial contracts. In comparison, the SCOAP$^3$ initiative has about the same number of partners, but a yearly budget of only 10M€, and will handle less than a dozen contracts with publishers. Therefore the aim is to operate SCOAP$^3$ along the blueprint of large HEP collaborations, to profit from their experience.

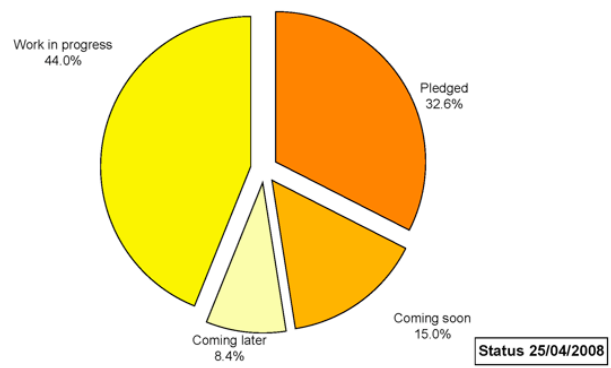

*Figure 3. Status of the SCOAP$^3$ fund-raising at the time of writing. About a third of the funds have already been pledged, 15% are expected to be pledged in the coming weeks, while discussions and negotiations are in progress for another 44%.*

SCOAP$^3$ is now collecting Expressions of Interest to join the consortium. Once it will have reached a critical mass, and thus demonstrated its legitimacy and credibility, it will issue a call for tender to publishers, aimed at assessing the exact cost of the operation, and then move quickly forward with the formal establishment of the consortium and its governance, then negotiating and placing contracts with publishers.

---

[15]http://www.consilium.europa.eu/ueDocs/cms_Data/docs/pressData/en/intm/97236.pdf

To date, most European countries have endorsed the project and major library consortia in the United States are in the process of completing the American share: SCOAP$^3$ has already received pledges for about a third of its budget envelope[16], with another third having the potential to be pledged in the short-term future, as presented in Figure 3.

## 6. Towards a future HEP Information System

For many years now almost all journal literature has been electronically available, the entire web is readily searchable, and commercial online databases provide metadata about all scientific literature. In addition, online services are changing more and more rapidly as new tools are developed and new ways of interacting with users evolve. In light of this fast-changing world, it is important to assess the usage by HEP researchers of the information resources that the community has pioneered in the last decades, as described in Section 3. Such an assessment serves two purposes: within the field, it informs on the need for HEP-specific community-based resources and their real role in the present internet landscape, inspiring their future evolution; globally, it provides an in-depth case study of the impact of discipline-based information resources, as opposed to institution-based information resources or cross-cutting (commercial) information platforms. This information is particularly relevant in light of recent worldwide moves towards self-archiving of research results at the institutional or disciplinary level and the need to effectively incorporate these resources in the research workflow.

This assessment was performed in mid 2007 through a user survey[17] that was filled by about 10% of the practitioners in the field: an overwhelming response whose results are discussed in the following. The main question of the survey concerned the most-used information systems in the field: for 91% of the participants these are services maintained by the community. The most popular systems are the SPIRES database with 48.2% and arXiv with 39.7%. Google scores 9% though within the group of scientists with less than two career years this fraction rises to 22%. The role of commercial databases is negligible with 0.1%. The results are depicted in Figure 4. It should be noted that the use of Google benefits strongly from the fact that community-based systems have made their content available for harvesting. At the same time, Google also acts, as in many other fields, as a broader alternative to publisher portals, given that indexing of many publisher websites has taken place in recent years.

Similar findings are observed for the most used systems to look for articles for which either the authors or the reference are known, as well as theses. Commercial systems never exceed a few percent share.

---

[16]The evolution of the SCOAP$^3$ fundraising and membership can be followed at: http://www.scoap3.org/fundraising.html and http://www.scoap3.org/whoisscoap3.html (last visited May 1$^{st}$, 2008).
[17]Anne Gentil-Beccot *et al.*, arXiv:cs.DL/0804.2701

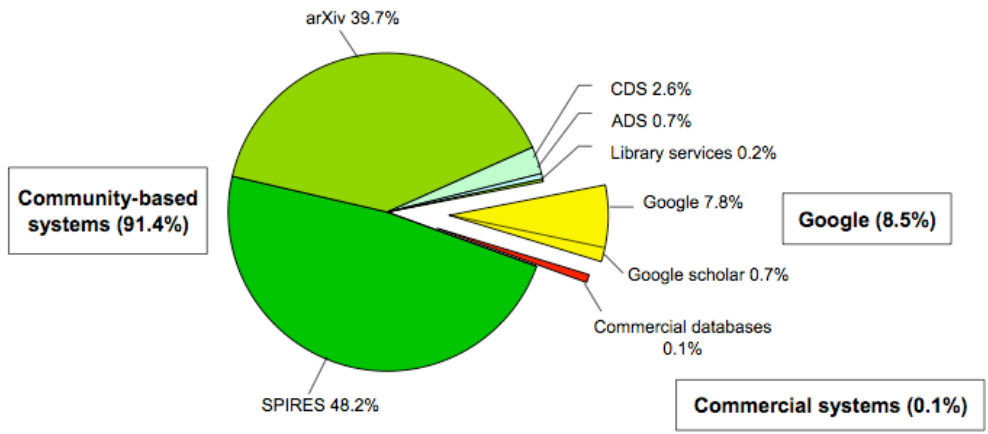

*Figure 4. Information resources favoured by HEP scientists. Community based systems dominate the landscape, even though among younger scholars there is an onset of Google. The usage of commercial systems (SCOPUS, INSPEC, the WebOfScience and similar products) is negligible.*

In addition to inquiring about the most heavily used systems for different tasks, the survey aimed to assess the importance of various aspects of information resources. Respondents were asked to tag the importance of 12 features of an information system on a five-step scale, ranging from "not important" to "very important". The results are presented in Figure 5. Access to full-text stood out clearly as the most valued feature, following close behind are depth of coverage, quality of content and search accuracy.

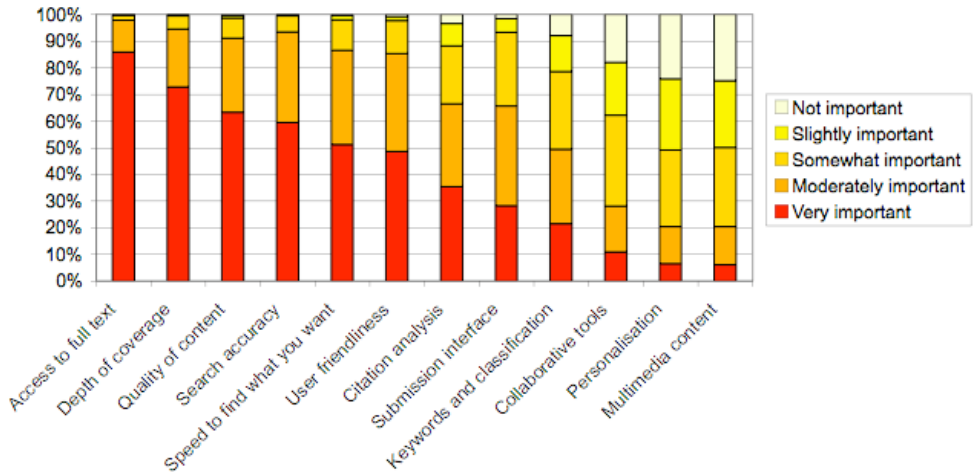

*Figure 5. Features of an information system most relevant for HEP scientists.*

The survey explicitly inquired about the level of change that HEP scholars would expect, and require, from their information resources in the next five years: 75% expected "some" to "a lot of" change and 90% of the users tagged three features as the most important areas of change: the linked presentation of all instances of a result, centralization, and access to data in figures and tables.

The survey also collected thousands of free-text answers, inquiring about features of current systems and their most-desired evolution. Some of the most inspiring free-text answers were along the following lines:

- Desire for seamless open access to older articles, prior to the onset of arXiv in the '90s.
- Improved full-text search and access to research notes of large experimental collaborations. These are a crucial grey-literature channel where large amounts of information and details about the results of large experiments transit.
- Indexing of conference talks and long-term archiving of the corresponding slides, beyond the lifetime of conference websites. Interlinking these slides with the corresponding conference proceedings, in preprint form with reference to published volumes, and possibly other instances describing the results.
- Use of the HEP information resources as fora for the publication of ancillary material, crucial in the research workflow, and in particular:
    - numerical data corresponding to tables;
    - numerical data corresponding to figures;
    - correlation matrices and additional information beyond these presented in tables to allow an effective re-use of scientific results;
    - fragments of computer code accompanying complex equations in articles, to improve the research workflow and reduce the possibility of errors;
    - primary research data in the form of higher-level objects.
- "Smarter" search tools, giving access to articles related to articles of interest.
- Establishment of some new sort of open peer-review, overlaid on arXiv.

The survey also tried to assess the potential for the implementation of Web2.0 features to capture user-tagged content. Respondents were asked how much time they would spend on a tagging system to give a service to the community: 63% would spend between five minutes a day and an hour a week. There is an immense potential for user-generated, or rather user-tagged and user-curated, content in the field of information provision in HEP, as in many other fields of web-based communication.

These results inform the future evolution of information management in HEP and, as these researchers are traditionally "early adopters" of innovation in scholarly communication, can inspire developments of disciplinary repositories serving other communities.

The results of this survey and strategic discussions between four leading HEP laboratories (CERN, DESY, Fermilab and SLAC), in synergy with other partners (notably arXiv) and in a continuous dialogue with major publishers in the field, led to a roadmap towards a future HEP information system, consisting of the following steps:

1. build a complete HEP information platform;
2. enable text- and data-mining applications;
3. demonstrate and deploy Web2.0 applications;
4. preserve research data and ensure their re-usability.

Work on step 1, the inception of the next generation of HEP information system, is in progress, blending the current SPIRES database with a modern platform, the Invenio[18] open-source digital-library software. This new information system, under the working name of Inspire, is being developed by a collaboration of four leading HEP laboratories: CERN, DESY, Fermilab and SLAC. It will integrate the content of present repositories and databases to host the entire body of metadata and the full-text of all open access publications, past and future, including conference material, and will embody the one-stop shop HEP researchers are waiting for, encompassing all content of arXiv as well as decades of previous articles. In addition, it will offer advanced tools for citation analysis, for example the "cited with" option, which often allows serendipitous discovery of related articles. The steps 2 and 3 have been charted, and will be further refined during 2008, leading to the ultimate creation of a next-generation repository for HEP. Interestingly, the technical solutions to be deployed are independent of the content, and therefore could be readily exported to other disciplines. Indeed, the Invenio software is now used in a dozen other repositories carrying content in other sciences and the humanities.

The next step on the roadmap for a future HEP information system will be to enable text- and data-mining applications that
- detect relations between documents carrying similar information;
- create datasets to exercise new hybrid metrics to measure the impact of articles, authors and groups;
- extract numerical information from figures and tables within published articles.

The mid-term future will see the development of Web2.0 applications that
- engage readers/authors in subject tagging, altering automatically-assigned classifications;
- allow community-based aggregation of related objects (articles, preprints, conferences, lectures);
- enable the possibility to review and comment on articles, adding links to additional documents or other digital objects.

It is interesting to note that the last features are already available in many services "overlaid" on arXiv, as a proto-form of alternative peer-review, but their acceptance is limited, due to the reduced usage of these sites when compared with the main access points to the literature. An inspiring experiment will be the deployment of these Web2.0 features in the production systems that the vast majority of HEP users adopts for their daily access to the literature: will this naturally lead to these additional means of communications entering the mainstream of the research workflow?

---

[18]Invenio was developed at CERN, where it now powers a 1-million records digital library ranging from articles to preprints, from multimedia to the institutional archives of the Organization, http://cdsware.cern.ch/invenio/index.html.

A long-term target will be, as a natural evolution of repositories, to link to data, simulations and computer programs behind each record, as discussed in Section 7.

## 7. The Next Frontier: Research Data

As discussed in the previous sections, HEP is a community of pioneers in scholarly communication. It would come as a surprise to an external observer, that few steps have been taken in opening up its data for re-use, a subject which is becoming more and more mainstream in the debate on the evolution of scholarly communication towards an "open society". An immediate explanation of this phenomenon is the sheer size and the complexity of HEP primary data, as well as the limited scope for their immediate interdisciplinary fruition, when compared, for instance, to data from astronomy or Earth observation. While the first is epitomized by the mental picture of one year of LHC data which, were it written on CDs, would result in a stack 20km tall, the latter gives space to more reflections. In recent years, powerful synergies arose between HEP and astrophysics, both studying our universe from different points of view, and simultaneous analyses of some data set will become an issue in the very near future. Beyond this opportunity, and based on first principles, HEP data should be preserved for their future re-analysis, being generated by a global enormous financial and human investment: they are a scientific legacy.

In HEP, the cases of a re-analysis of older data are only few but important. An example in which the first author has been involved as member of two large experimental collaborations is the re-analysis of data collected in the '80s at DESY together with data collected in the late '90s at CERN, in the light of improved theories[19]. Without going into technical details, this work allowed to improve the understanding of the force which binds quarks in the nuclei. Only by chance this combined analysis was possible since researchers from JADE had moved to OPAL and kept tapes of data and the corresponding software[20].

Beyond anecdotes, there is a growing awareness of the need to preserve HEP data for their re-use and for (open) access[21]. There are five distinct continua for which data should be preserved or for which access would enable the production of more science, corresponding to very different user profiles and time scales after the data collection:
1. the same researchers who took the data, after the closure of the facility, with a time scale which can be a year or a decade;
2. researchers working at similar experiments, with a time scale between days and years;
3. researchers at future experiments, with a time scale of several decades;
4. theoretical physicists who may want to re-interpret the data, with a time scale of months or some years;
5. theoretical physicists who may want to test future theories, with a time scale of decades.

---

[19] The JADE and OPAL collaborations, Eur.Phys.J.C17 (2000) 19, hep-ex/0001055
[20] To continue the story they bought a juke-box to store CDs of OPAL data after the completion of this later experiment.
[21] S. Mele, "Preservation, re-use and (open) access to HEP data" contributed to Tools & Trends in Digital Preservation, The Hagues, 31 October 2007;
J. Engelen, presentation at the Conference of the Alliance for Permanent Access, Brussels, 15 November 2007, http://www.alliancepermanentaccess.eu/power/Engelen_Alliance_151107.ppt

These cases are all different. In (1), the knowledge on the content of the data is in principle available, but software and hardware problems might be the limiting factor. In (2) and (4) a synergy between the user and the producer of data could lead to an immediate re-use of data, without the need to devise preservation strategy. However, (3) and (5), which can imply crucial consequences for the advancement of the field, make it evident that knowledge, in addition to technical solutions, has to be preserved together with the data. Enabling all these continua to re-use HEP data implies an indissoluble link between preservation strategies and access strategies. Indeed, preservation, in HEP, is not entirely a technical or an archival issue: during the long life-time of experiments, sometimes two decades, computing centres routinely copy old tapes onto new facilities and software migration can and does occur. This is made possible by a curation of the data by the producer themselves, which implies that HEP data from facilities recently stopped or about to be discontinued remains vaguely readable, although only understandable to the scientists who produced them in the first place. With the disbanding of the experimental collaborations that collected the data, though, software migrations stop and, worse, the insider knowledge needed for a re-analysis at a later stage is scattered or lost.

In a sentence, the issue with the preservation, re-use and (open) access to HEP data is the complexity of the data themselves.

A prerequisite for the re-use of HEP data is to embed in the data the knowledge about the data themselves, creating higher-level objects which can be understood later in time and by scientists not involved in the creation of the data. This "parallel" data format would have to emerge in addition to the ones used internally by the experiments. The benefits of such model are obvious. But there are formidable obstacles to overcome, technical and sociological. From a technical point of view, these "parallel" high-level data should be defined, possibly standardized, and created at the same time as the standard data reconstruction[22]. Even if this process would only cost a fraction of the human and financial capital invested in HEP experiments, a small fraction of a large number (thousands of person-years) still translates in a major project. This leads to the sociological barriers: this further investment would be in competition with research, and there is little to no academic incentive to put data preservation higher on the agenda than the data perusal and the preparation for further experiments. Scientists would need enormous academic incentives or additional funds. A debate[23] should take place concerning data ownership, (open) access, credit, accountability, reproducibility of results, depth of peer-reviewing. While each of these issues would require a full-length article to be laid out, it is clear that decades of a traditional and monolithic way of doing research need rethinking. Due to the complexities of these issues, HEP may be considered as a worst-case scenario in the topic of data preservation, re-use and (open) access, but a scenario that has the potential to inspire other fields of science, as in the other endeavours of HEP in the field of scholarly communication.

---

[22]S. Mele, "Preservation, re-use and (open) access to HEP data" contributed to Tools & Trends in Digital Preservation, The Hagues, 31 October 2007;
J. Engelen, presentation at the Conference of the Alliance for Permanent Access, Brussels, 15 November 2007,
http://www.alliancepermanentaccess.eu/power/Engelen_Alliance_151107.ppt.
[23]The FP7 PARSE.Insight project (Insight in Permanent Access to the Records of SciencE) has among its objectives to understand the implications, not only technical, for HEP to start a process of preserving its data. PARSE.Insight will deliver its report in 2010. http://parse.digitalpreservation.eu/.

## 8. Conclusions

With 50 years of preprints and 17 years of repositories, not to mention the invention of the web, HEP has spearheaded (open) access to scientific information and is now in a period of change at two frontiers: the cross road of open access and peer-reviewed literature and the inception of a next-generation repository which has to adapt the current technological advances to the research workflow of HEP scientists.

In the spirit of their collaborative tradition, HEP scientists are now proposing to pool together resources from libraries and HEP institutes worldwide to sponsor the transition to open access of the entire literature of the field, through the SCOAP$^3$ initiative (Sponsoring Consortium for Open Access Publishing in Particle Physics). This open access publishing model is gathering growing international consensus, being non-disruptive to authors and, to a substantial degree, to publishers and societies. It has the potential to fundamentally alter the role of libraries in the publishing process and re-think the role of high-quality journals in the open access era. SCOAP$^3$ should demonstrate its potential in the coming months.

At the same time a new e-infrastructure of HEP Scientific Communication is being set up. A complete information platform is being built, enabling text- and data-mining as well as Web2.0 applications. This new platform is entirely user-pulled, building on a tradition of authors strongly supporting and advocating the use of repositories and meeting the expectations of HEP physicists in light of the new opportunities offered by technological development to scholarly communication. The new e-infrastructure might provide inspiration to many other communities which are currently exploring ways to improve the dissemination, discovery and organization of research results, primarily focusing on author self-archiving.

A new challenge for the future, which has however to be tackled as soon as possible, is the preservation, re-use and (open) access to HEP data. This is both a technical issue and a sociological one. On one side sits the sheer amount and relative complexity of HEP data, on the other a culture that surprisingly did not foster efforts to disseminate and exchange HEP data. Awareness on this issue is rising, and the next years, together with new, landmark, HEP data from the Large Hadron Collider at CERN, might bring new ways to preserve and re-use its data, another example for other scientific disciplines.

We are at the onset of a new era of innovation in scholarly communication in HEP and beyond.